\documentclass[conference]{IEEEtran}
\IEEEoverridecommandlockouts
\usepackage{cite}
\usepackage{graphicx} % Required for inserting images
\usepackage{xurl}
\usepackage{amsmath,amssymb,amsfonts}
\usepackage{textcomp}
\usepackage{hyperref}
\usepackage{listings}
\usepackage{xcolor}
\usepackage{tcolorbox}
\usepackage{flushend}
\usepackage{booktabs}
\def\BibTeX{{\rm B\kern-.05em{\sc i\kern-.025em b}\kern-.08em
    T\kern-.1667em\lower.7ex\hbox{E}\kern-.125emX}}

\begin{document}
\title{Integrating APK Image and Text Data for Enhanced Threat Detection: A Multimodal Deep Learning Approach to Android Malware}

 \author{\IEEEauthorblockN{Md Mashrur Arifin}
 \IEEEauthorblockA{\textit{Department of Computer Science} \\
 \textit{Boise State University}\\
 Boise, ID, USA \\
 mdmashrurarifin@u.boisestate.edu}
 \and
 \IEEEauthorblockN{Maqsudur Rahman}
 \IEEEauthorblockA{\textit{Department of Computer Science} \\
 \textit{Boise State University}\\
 Boise, ID, USA \\
 maqsudurrahman@u.boisestate.edu}
 \and
 \IEEEauthorblockN{Nasir U. Eisty}
 \IEEEauthorblockA{\textit{Department of EECS} \\
 \textit{The University of Tennessee}\\
 Knoxville, TN, USA \\
 neisty@utk.edu}
 }
\date{January 2026}

\maketitle

\begin{abstract}
% Zero-day Android malware attacks have become increasingly sophisticated, necessitating the development of novel approaches capable of identifying previously unseen threats. Recent research has demonstrated the effectiveness of image-based feature representations of malware bytecode in conjunction with Convolutional Neural Networks (CNNs) for zero-day malware classification tasks. 
% However, existing studies frequently lack a systematic investigation into the impact of image type and resolution on detection performance.  Furthermore, these detection approaches often overlook the potential textual information embedded within Android Application Packages (APKs) that can contribute in malware detection, such as permissions and metadata, etc., limiting their ability to capture the full context of malicious behavior.
As zero-day Android malware attacks grow more sophisticated, recent research highlights the effectiveness of using image-based representations of malware bytecode to detect previously unseen threats.
However, existing studies often overlook how image type and resolution affect detection and ignore valuable textual data in Android Application Packages (APKs), such as permissions and metadata, limiting their ability to fully capture malicious behavior.
The integration of multimodality, which combines image and text data, has gained momentum as a promising approach to address these limitations. 
This paper proposes a multimodal deep learning framework integrating APK images and textual features to enhance Android malware detection. We systematically evaluate various image types and resolutions across different Convolutional Neural Networks (CNN) architectures, including VGG, ResNet-152, MobileNet, DenseNet, EfficientNet-B4, and use LLaMA-2, a large language model, to extract and annotate textual features for improved analysis. The findings demonstrate that RGB images at higher resolutions (e.g., 256x256, 512x512) achieve superior classification performance, while the multimodal integration of image and text using the CLIP model reveals limited potential.
%, despite being constrained by the small dataset. 
Overall, this research highlights the importance of systematically evaluating image attributes and integrating multimodal data to develop effective malware detection for Android systems.

\end{abstract}


\section{Introduction}
%Consider the implications of an undetectable threat compromising your smartphone—an essential tool for communication, finance, and daily organization. 
%Such scenarios are becoming increasingly common as the security of mobile applications \cite{qureshi2024apk} especially on Android systems emerges as a critical concern.
Malware targeting the Android ecosystem is becoming increasingly common, highlighting the urgent need for robust security measures to protect users and sensitive data~\cite{qureshi2024apk}.
This threat extends beyond individual users to governments and corporations~\cite{abdul2024mobile}  since sophisticated zero-day malware is evolving at an alarming pace. Malware detection conventionally relies on static or dynamic analysis, and is often vulnerable to obfuscation techniques, which let intruders conceal the real purpose of their programs~\cite{siddiqui2024overview}. This limitation has increased the demand for developing strategies that go beyond traditional techniques and offer novel perspectives on threat identification.

The investigation of image-based deep learning methods for malware detection has gained momentum in recent years~\cite{liu2024efficient}. When viewed as images, APK files highlight specific structural and pattern-based traits that separate dangerous applications from the safe ones. %Though this method is promising, it is still in its early stages. The detection performance greatly depends on the attributes of image representation used, such as RGB or grayscale. Significant gaps remain in understanding the optimal characteristics of images used for such detection systems. 
Though promising, this method is still early-stage. Detection performance heavily depends on image attributes like RGB or grayscale, and the optimal image characteristics remain unclear.
Specifically, the effectiveness of different types of images, such as RGB and grayscale images, and their respective resolutions (e.g., 128x128, 256x256, and 512x512) has not been comprehensively evaluated. This lack of clarity hinders the development of robust image-based malware detection systems, as varying image formats and resolutions can substantially impact model performance.

Moreover, deep learning models only depends upon image based technique often ignore the valuable insights that may be obtained from textual elements found in the manifest files, permissions, and metadata of APKs. The integration of textual data extracted from APK files with image data through multimodal models presents a promising avenue for enhancing malware detection capabilities. While several studies have leveraged language models for textual analysis~\cite{zhao2025apppoet}, the synergistic potential of combining image and text modalities remains underexplored. 
%Initial findings suggest that multimodal approaches may offer improved performance over traditional single-modality systems; however, 
Therefore, empirical investigations are needed to assess whether this integration truly enhances or detracts from malware detection accuracy.
To investigate this gap, we formulate two research questions:
\begin{description}
    \item [\textbf{RQ1:}] How do different image processing techniques and resolutions (e.g., RGB vs. grayscale) impact the effectiveness of distinguishing between benign and malicious Android applications?

    \item [\textbf{RQ2:}] How does integrating image and textual data in a multimodal framework affect the performance of Android malware detection systems?

\end{description}

%\paragraph{Our Contribution} In our work, we have developed and assessed a novel prompt and multimodal based  approach that integrates both image and text data. Our research has enhanced Android application security by refining malware detection techniques, hence building important insights and recommendations  significant for detecting malicious android  applications. The following are the contribution of our research:
We developed and evaluated a novel prompt-based multimodal approach that integrates both image and text data, offering key insights and practical recommendations for detecting malicious Android applications. Our main contributions are:

\begin{itemize}
    \item Analyzed RGB and Greyscale images at three resolutions (e.g., 128x128,
256x256, 512x512) for Android malware detection, using eight CNN architectures to identify the optimal image attributes. Our results show that RGB images with 256x256 or 512x512 resolutions have balanced accuracy and computational efficiency.

    \item Developed a novel framework combining APK images and prompt-generated text annotations, integrating image and text features using the CLIP model. However, CLIP did not outperform standalone image-based models, with ResNet proving more effective overall. 
    
    \item Provided a limited multimodal dataset comprising diverse attributed images and their corresponding textual annotations, supporting the ongoing enhancement of this dataset.
 
\end{itemize}

% The data and materials related to this paper are available at the following link:
% \href{https://github.com/arifinmash/mmandmal}{Data}
\section{Related Works}
\textit{Image-Based Malware Detection.}
An early study introduced the MADRF-CNN model, which converted DEX files into RGB images and achieved 96.9\% identification accuracy \cite{zhu2023effective}, while another study also reported high detection rates using CNNs on color images without relying on textual data \cite{vasan2020imcfn}. These studies have shown the effectiveness of CNN-based models for detecting Android malware by converting APK files into images, but they often overlook a systematic evaluation of how image variations like color format and resolution affect detection accuracy.

\textit{Text-Based Malware Detection.}
Textual features such as permissions, API calls, and system logs have proven valuable for malware detection, with models such as HYDRA achieving over 99\% accuracy by combining these elements with image features~\cite{gibert2020hydra}, and hybrid models using GRU and DBN also showing strong performance~\cite{lu2020android}. While text-based approaches are effective, they often miss visually distinct patterns that image-based analysis can identify, underscoring the importance of integrating textual and visual data for comprehensive malware detection.

\textit{Multimodal Approaches Integrating Image and Text Data.}
Multimodal approaches combine text and image data and promise to improve malware detection~\cite{8443370}. Ullah et al.~\cite{ullah2022malware} reached 99.27\% accuracy by integrating image features with control flow graph-based API analysis. Zhang et al.~\cite{zhang2023detection} achieved 98.9\% detection on encrypted traffic using CNNs and deep forest classifiers. These methods enhance detection accuracy and resilience to obfuscation by fusing visual and semantic features.
The literature highlights the need to evaluate image attributes in CNNs and combine text and image-based methods for detecting advanced Android malware. This observation supports our study of using multimodal data to improve detection performance and address challenges like obfuscation and resource constraints.

% \section{Preliminaries}
% \input{sec_preliminaries}
\section{Research Approach}
%This section summarizes our research approach, including the dataset collection process, the investigation of key image attributes, and the evaluation of the effectiveness of a multimodal approach.

\subsection{Data Collection}
We collected data on malware and benign APKs from two dataset sources: CIC-AndMal2017~\cite{ExplainableModel2023} and CICMalDroid 2020~\cite{TreeBasedEnsemble2022}. The collected malware includes Adware, Scareware, Banking, and Smsmalware, alongside benign APKs. In total, we collect 4303 malware APKs and 4039 benign APKs.

% \begin{figure}[h]
%     \centering
%     \includegraphics[width=0.75\linewidth]{resources/data.png}
%     \caption{Data Collection}
%     \label{fig:data collection}
% \end{figure}

\subsection{Investigation for Finding the Effective Image Attributes}
\begin{figure}[h]
    \centering
    \includegraphics[width=0.75\linewidth]{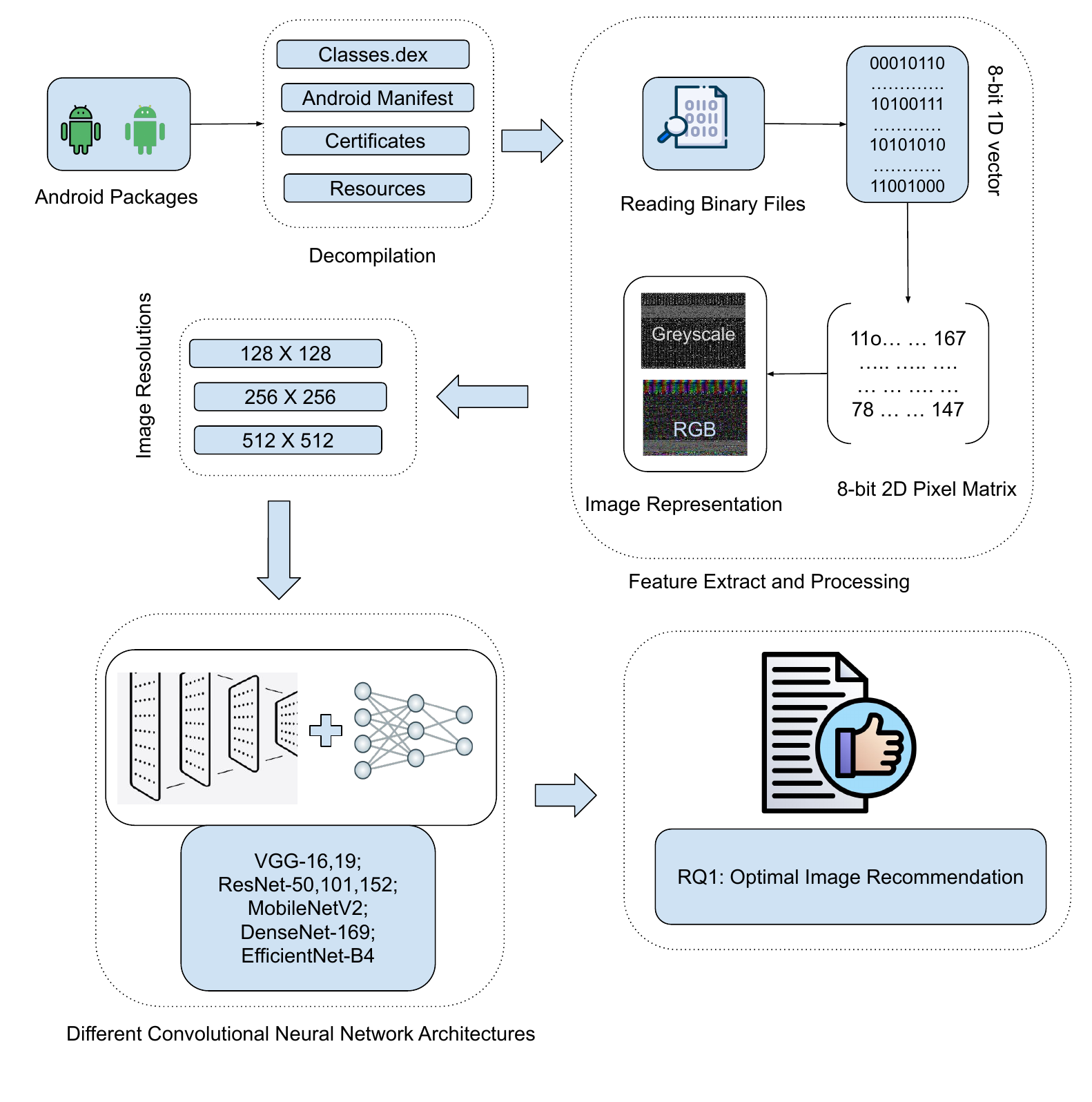}
    \caption{Investigation for finding the effective Image attributes}
    \label{fig:method_1}
\end{figure}

To find the answer for  \textbf{RQ1}, we start by transforming APK Files into two image representations, such as grayscale and RGB, as shown in Fig.~\ref{fig:method_1}. Both image representations were produced at several resolutions: 128x128, 256x256, and 512x512 pixels. Upon completion of the image conversions, we conducted a performance study by training and testing eight CNN models~\cite{CNNDroid2018}. The examined models were VGG-16, VGG-19~\cite{simonyan2014very}, ResNet-50, 101, 152~\cite{he2016deep}, MobileNet-V2~\cite{howard2017mobilenets}, DenseNet-169~\cite{huang2017densely}, and EfficientNet-B4~\cite{tan2019efficientnet}. Each model has unique advantages. VGG models are simple and function as benchmarks, while ResNet models identify complex patterns via deep architectures. MobileNet-V2 is a lightweight and efficient model, ideal for resource-limited settings, whereas DenseNet169 demonstrates superior feature reuse and efficiency. EfficientNet-B4 optimizes accuracy while minimizing computing expense. Collectively, these models provide a comprehensive evaluation of accuracy, efficiency, and scalability, guaranteeing reliable outcomes across various image resolutions and computing contexts. The CNN architectures were trained on the produced image datasets to evaluate their capacity to distinguish between benign and malicious APKs.

To assess model performance, we used metrics such as accuracy, precision, recall, F1-score, and area under the ROC curve (AUC-ROC). These measures offered significant insights into the classification efficacy of each model and enabled us to assess its resilience in identifying diverse malware variants. 

\subsection{Investigation for Multimodal Approach}
\begin{figure}[h]
    \centering
    \includegraphics[width=1\linewidth]{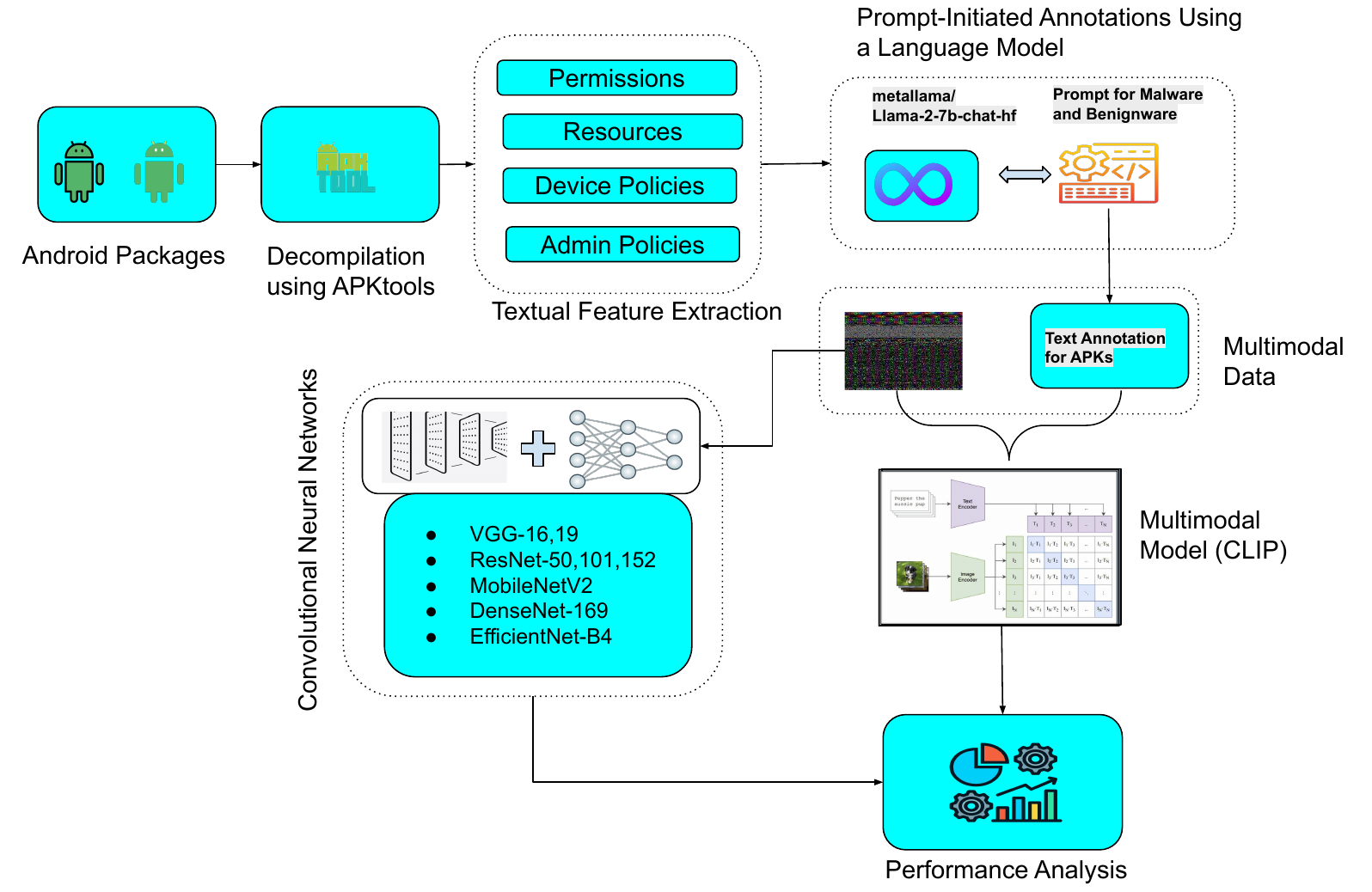}
    \caption{Investigation for Multimodal Approach}
    \label{fig:Method_2}
\end{figure}

To answer \textbf{RQ2}, we start extracting text from APKs (as shown in Fig.~\ref{fig:Method_2}), with a special focus on permissions, resources, and the manifest.xml file. These collected texts form the foundation for designing prompts highlighting essential features relevant to malware and benignware identification. The prompts are further analyzed by a Large Language Model (LLM), Llama 2, which examines the features of retrieved text as either malicious or benign according to the input prompts~\cite{LI2025102662}. This annotation technique enhances the dataset by offering contextual insights into application behavior.
After the text annotation, we provide multimodal data by integrating the annotated text with images generated from the APKs in the preceding step. The resultant multimodal dataset includes textual and visual data, providing a more comprehensive foundation for our research. Subsequently, we assess this multimodal data using the CLIP model~\cite{8585344}, which has been utilized as a nearly zero-shot operation to classify the integrated characteristics of text and images. 
%This method enables a more comprehensive identification of malware than conventional image-only detection techniques. 
We used assessment measures, including accuracy, precision, recall, F1-score, and area under the ROC curve (AUC-ROC), to evaluate the multimodal models' performance relative to the other CNN architectures and Quantum Neural Networks.

\section{Experiments and Results}
% This section details the experimental design and the results obtained from evaluating different image configurations and models. It covers the process of training CNN models on APK images and analyzing their performance. The section also discusses the findings related to the effectiveness of various image representations and multimodal integration, providing insights into the best configurations for Android malware detection.

\subsection{\textbf{RQ1 - Assessing the Optimal Image Representations for Android Malware Detection}}
\subsubsection{\textbf{Procedure}}
We built six sets (three resolutions across two types) of image representations extracted from APKs. Images are standardized to a consistent resolution to maintain uniformity. During the experiment, each dataset is divided into training, validation, and testing subsets, with 80\% allocated for training and the remaining parts designated for validation and testing.  We used all the sets of data for 8 CNN-based pre-trained deep learning models (VGG-16, VGG-19, ResNet-50, 101, 152, MobileNet-V2, DenseNet-169, and EfficientNet-B4 ) to assess their capacity to categorize Android files as benign or malicious. Models are trained with a batch size of 32 for a maximum of 20 epochs, using early stopping (patience = 5) if validation performance fails to improve after five consecutive epochs. This training approach tracks the model performance improvement and reduces overfitting. The Adam optimizer reduces the loss function while learning the optimum model weights. Training is performed using a GPU, enhancing computational speed and minimizing training duration. 

% \subsubsection{\textbf{Evaluation Metrics}}
% This process of determining the optimal image resolution can be defined as a performance-based evaluation strategy, in which the model's key performance metrics—such as accuracy, precision, recall, F1-score, and ROC AUC—are optimized to make the decision. Considering both model efficiency and computational costs, the procedure entails assessing how each resolution impacts the model's capacity to generalize and attain high predictive performance. The optimal image resolution is determined based on how it impacts the overall performance of the model. 

\subsubsection{\textbf{Results}}
%The experimental results provide a thorough examination of the best image setups for classifying Android malware using several CNN architectures. 
\begin{table*}[ht]
\centering
\caption{Comparative classification Results across Various CNN architectures}
\begin{tabular}{@{}lccccccccc@{}}
\toprule
\textbf{Model}          & \textbf{Configuration} & \textbf{Resolution} & \textbf{Accuracy} & \textbf{Precision} & \textbf{Recall} & \textbf{F1-Score} & \textbf{ROC AUC} \\ \midrule
\textbf{VGG-16}         & \textbf{Grayscale}     & 128x128             & 58\%              & 0.64               & 0.59             & 0.55              & \textbf{0.768}   \\
                       &                        & 256x256             & 52\%              & 0.26               & 0.50             & 0.34              & 0.50             \\
                       &                        & 512x512             & 52\%              & 0.26               & 0.50             & 0.34              & 0.50             \\
                       & \textbf{RGB}           & 128x128             & 52\%              & 0.26               & 0.50             & 0.34              & 0.50             \\
                       &                        & 256x256             & 51\%              & 0.26               & 0.50             & 0.34              & 0.50             \\
                       &                        & 512x512             & 48\%              & 0.24               & 0.50             & 0.33              & 0.50             \\ \midrule
\textbf{VGG-19}         & \textbf{Greyscale}     & 128x128             & 51\%              & 0.26               & 0.50             & 0.34              & 0.50             \\
                       &                        & 256x256             & 52\%              & 0.26               & 0.50             & 0.34              & 0.50             \\
                       &                        & 512x512             & 52\%              & 0.26               & 0.50             & 0.34              & 0.50             \\
                       & \textbf{RGB}           & 128x128             & 52\%              & 0.26               & 0.50             & 0.34              & 0.50             \\
                       &                        & 256x256             & 51\%              & 0.26               & 0.50             & 0.34              & \textbf{0.773}   \\
                       &                        & 512x512             & 48\%              & 0.24               & 0.50             & 0.33              & 0.50             \\ \midrule
\textbf{ResNet50}       & \textbf{Grayscale}     & 128x128             & 94\%              & 0.94               & 0.94             & 0.94              & 0.9798           \\
                       &                        & 256x256             & 95\%              & 0.95               & 0.95             & 0.95              & 0.9897           \\
                       &                        & 512x512             & 93\%              & 0.93               & 0.93             & 0.93              & 0.9799           \\
                       & \textbf{RGB}           & 128x128             & 95\%              & 0.95               & 0.95             & 0.95              & 0.9859           \\
                       &                        & 256x256             & 96\%              & 0.96               & 0.96             & 0.96              & 0.9902           \\
                       &                        & 512x512             & 96\%              & 0.96               & 0.96             & 0.96              & 0.9906           \\ \midrule
\textbf{ResNet101}      & \textbf{Greyscale}     & 128x128             & 0.95              & 0.95               & 0.95             & 0.95              & 0.9854           \\
                       &                        & 256x256             & 0.94              & 0.94               & 0.94             & 0.94              & 0.9865           \\
                       &                        & 512x512             & 0.94              & 0.94               & 0.94             & 0.94              & 0.9845           \\
                       & \textbf{RGB}           & 128x128             & 0.95              & 0.95               & 0.95             & 0.95              & 0.9861           \\
                       &                        & 256x256             & 0.96              & 0.96               & 0.96             & 0.96              & 0.9889           \\
                       &                        & 512x512             & 0.97              & 0.97               & 0.97             & 0.97              & 0.9919           \\ \midrule
\textbf{ResNet152}      & \textbf{Grayscale}     & 128x128             & 94\%              & 0.94               & 0.94             & 0.94              & 0.9780           \\
                       &                        & 256x256             & 96\%              & 0.96               & 0.96             & 0.96              & 0.9910           \\
                       &                        & 512x512             & 95\%              & 0.95               & 0.95             & 0.95              & 0.9870           \\
                       & \textbf{RGB}           & 128x128             & 95\%              & 0.95               & 0.95             & 0.95              & 0.9870           \\
                       &                        & 256x256             & 96\%              & 0.96               & 0.96             & 0.96              & 0.9890           \\
                       &                        & 512x512             & 96\%              & 0.96               & 0.96             & 0.96              & 0.9920           \\ \midrule
\textbf{MobileNetV2}    & \textbf{Grayscale}     & 128x128             & 95\%              & 0.95               & 0.95             & 0.95              & 0.9767           \\
                       &                        & 256x256             & 95\%              & 0.95               & 0.95             & 0.95              & 0.9860           \\
                       &                        & 512x512             & 96\%              & 0.96               & 0.96             & 0.96              & 0.9898           \\
                       & \textbf{RGB}           & 128x128             & 94\%              & 0.94               & 0.94             & 0.94              & 0.9822           \\
                       &                        & 256x256             & 95\%              & 0.95               & 0.95             & 0.95              & 0.9885           \\
                       &                        & 512x512             & 96\%              & 0.96               & 0.96             & 0.96              & 0.9903           \\ \midrule
\textbf{DenseNet169}    & \textbf{Greyscale}     & 128x128             & 94\%              & 0.96               & 0.91             & 0.94              & 0.9756           \\
                       &                        & 256x256             & 94\%              & 0.96               & 0.91             & 0.94              & 0.9756           \\
                       &                        & 512x512             & 95\%              & 0.95               & 0.95             & 0.95              & 0.9895           \\
                       & \textbf{RGB}           & 128x128             & 95\%              & 0.97               & 0.91             & 0.94              & 0.9861           \\
                       &                        & 256x256             & 96\%              & 0.96               & 0.96             & 0.96              & 0.9882           \\
                       &                        & 512x512             & 95\%              & 0.95               & 0.95             & 0.95              & 0.9871           \\ \midrule
\textbf{EfficientNet-B4}& \textbf{Grayscale}     & 128x128             & 94\%              & 0.94               & 0.93             & 0.94              & 0.9777           \\
                       &                        & 256x256             & 97\%              & 0.97               & 0.97             & 0.97              & 0.9924           \\
                       &                        & 512x512             & 96\%              & 0.96               & 0.96             & 0.96              & 0.9932           \\
                       & \textbf{RGB}           & 128x128             & 95\%              & 0.95               & 0.94             & 0.95              & 0.9853           \\
                       &                        & 256x256             & 96\%              & 0.97               & 0.95             & 0.96              & 0.9950           \\
                       &                        & 512x512             & 97\%              & 0.98               & 0.97             & 0.97              & 0.9949           \\ \bottomrule
\end{tabular}
\label{tab:combined_results}
\end{table*}

As shown in Table~\ref{tab:combined_results}, for Model VGG-16, the Grayscale 128x128 configuration proved to be the most successful, with 58\% accuracy, an F1-score of 0.55, and a ROC AUC of 0.768. The effectiveness of lower grayscale inputs was shown by the reduction in performance and misclassification problems caused by higher resolutions and RGB combinations. However, for Model VGG-19, the 256x256 RGB arrangement performed the best with a ROC AUC of 0.773. This design successfully balanced computational cost and performance, even though accuracy, precision, and recall scores were comparable among configurations. The suggested resolution is 256x256 RGB since higher resolutions, such as 512x512, did not provide noticeable improvements,  because of noise or overfitting. The 128x128 grayscale resolution worked better for VGG-16 because the grayscale format removed unnecessary complexity, and the simplified model was able to learn important features well without overfitting to noise. The 256x256 RGB resolution for VGG-19 offered an adequate compromise between color information and detail, allowing a more complex model to capture subtler features and enhance performance without the overfitting or computational load that comes with larger resolutions.

In case of ResNet-50, RGB images of 512x512 resolution yielded the highest results, with 96\% accuracy and a ROC AUC of 0.9906. The RGB 256x256 setup, on the other hand, was an effective substitute as it provided good performance (96\% accuracy, 0.9992 ROC AUC) with less processing overhead. In general, grayscale setups performed worse, particularly at higher resolutions. For Model ResNet-101, RGB images consistently performed better than grayscale images with ResNet-101 at all resolutions. The best option was the 512x512 RGB arrangement, which had the greatest accuracy (97\%), precision (97\%), and ROC AUC (0.9919). The 256x256 RGB configuration was a good choice for situations with limited resources since it produced results that were equivalent while requiring less processing power. For ResNet-152, the RGB 512x512 setup produced the greatest results, with a ROC AUC of 0.992 and 96\% accuracy. While grayscale choices performed similarly, they were slightly behind RGB.

MobileNetV2 performed well in both RGB and grayscale settings, reaching a 96\% accuracy peak for 512x512 images. The RGB 512x512 configuration is the recommended option for maximum accuracy since it has the greatest ROC AUC of 0.9903. The 256x256 resolution provided a good compromise between reduced computational cost and little performance deterioration. For the DenseNet-169 Model, the accuracy ranged from 94\% to 96\%, demonstrating its resilience across setups. Although it needed more processing power, the 512x512 RGB setup produced the best results. With a ROC AUC of 0.9882 and 96\% accuracy, the 256x256 RGB configuration offered a balanced approach that can be ideal for real-world uses. For  EfficientNet-B4, with 512x512 RGB inputs,  performed very well at higher resolutions, with 97\% accuracy. At 256x256 resolution, grayscale setups worked well because they provided a compromise between computing efficiency and accuracy. While 256x256 is a resource-conscious option, the 512x512 RGB configuration is advised for optimal performance.

Overall, Models such as ResNet-50, ResNet-101, and  ResNet-152 performed best with 512x512 RGB resolution because it enabled these deeper architectures to fully utilize the increased detail and color information, resulting in higher accuracy and ROC AUC scores, although at the expense of a higher computational cost. However, the 256x256 RGB configuration often offered similar results with less processing cost, ideal for those looking for a compromise between performance and computational cost.

\subsection{\textbf{RQ2 - Evaluating Multimodal Performance for Android Malware Detection}}
This experiment consists of two tasks: prompt design and training \& evaluating the CLIP multimodal model. 
\subsubsection{\textbf{Procedure 1 -- Prompt Design}}
\begin{figure}[h]
    \centering
    \includegraphics[width=1\linewidth]{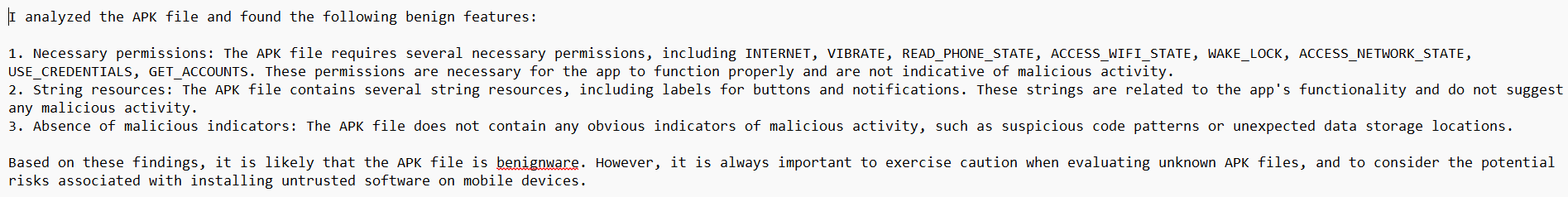}
    \caption{Prompt Response for Benign APK }
    \label{fig:prompt benign}
\end{figure}

\begin{figure}[h]
    \centering
    \includegraphics[width=1\linewidth]{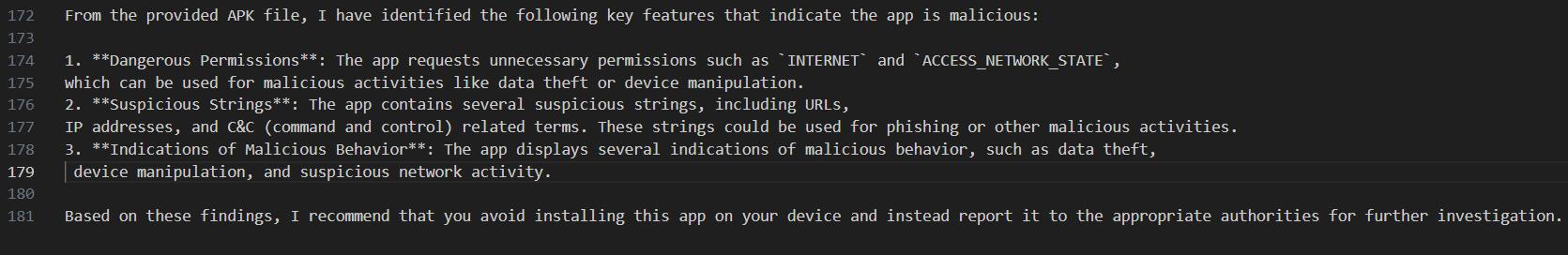}
    \caption{Prompt Response for malware APK}
    \label{fig:prompt malware}
\end{figure}

To annotate the APK images, we designed a prompt-based summarization method that utilizes the extracted text from the APKs to identify characteristics indicative of both malicious and benign behavior. To generate concise summaries against the two distinct designed prompts (each for malicious and benign), we utilized the extracted text, the LLaMA-2 language model~\cite{touvron2023llama}, and its tokenizer. The whole process was implemented in three simple steps, i.e, tokenizing the input within certain bounds, providing an instructive and focused prompt, and creating answers using the pre-trained model with regulated parameters for output variety. In generating this annotation, we explicitly queried regarding the permissions, strings of the application operation, and the presence/absence of malicious signs in our designed prompts. 
%Both qualitative and quantitative evaluations for the generated texts were conducted. To do this, we drew on the experience of one of the paper's authors, who confirmed the results and offered human input. 
We manually conducted qualitative and quantitative evaluations of the generated texts to confirm their validity.
This thorough assessment made sure the summaries were accurate and realistically applicable to the analysis of APK texts (shown in Fig.~\ref{fig:prompt benign} and Fig~\ref{fig:prompt malware} ).

\begin{tcolorbox}[width=\columnwidth, colback=green!5, colframe=green!80, coltitle=green!90, sharp corners=northwest, boxrule=0.8mm, fonttitle=\bfseries\large, toptitle=1mm, bottomtitle=1mm, title=Prompt for Benignware APKs]
    \textbf{Prompt for Benign APKs:} \textit{Examine the provided text extracts from the APK file. In a single paragraph, identify and summarize the key features that indicate the APK is benignware. Focus on necessary permissions, strings related to app functionality, and the absence of malicious indicators. Provide a concise and informative summary, prioritizing the most important and relevant features.
}
\end{tcolorbox}

\begin{tcolorbox}[width=\columnwidth, colback=red!5, colframe=red!80, coltitle=red!90, sharp corners=northwest, boxrule=0.8mm, fonttitle=\bfseries\large, toptitle=1mm, bottomtitle=1mm, title=Prompt for malware APKs]
    \textbf{\textbf{Prompt for malware APKs:}}  \textit{Examine the provided text extracts from the APK file. In a single paragraph, identify and summarize the key features that indicate the APK is malware. Focus on dangerous permissions (e.g., 'SEND\_SMS', 'READ\_CONTACTS'), suspicious strings (URLs, IP addresses, C\&C related terms), and indications of malicious behavior (data theft, device manipulation).}
\end{tcolorbox}

\subsubsection{\textbf{Procedure 2 -- Multimodal Model Training and Evaluation}}
We started this process by initializing the Hugging Face Transformers library's CLIP model and processor~\cite{huggingfaceCLIP}. Since the model has already been trained on a variety of datasets, we did not fine-tune for our experiment. After that, we concentrated on data preparation by importing images from a specified multimodal format together with the text descriptions (we used a limited 34 images and 34 text annotation data). We analyzed the images using the PIL library, and we received text descriptions from .txt files that correspond to the image names. For Malware and Benignware, we separated the data into two separate datasets, which we merged into a single list of images, text, and labels for further processing.

We used the CLIP processor to preprocess each image-text combination before making predictions. Tokenizing the text and preparing the inputs for model compatibility are part of this stage. We ran the processed inputs through the CLIP model to calculate similarity scores (logits) between the textual and visual inputs. We determined the probabilities for each category by applying a softmax function to the logits, and we chose the category with the greatest probability to be the predicted label. Throughout this procedure, we observed the predicted categories and the associated real labels for subsequent analysis.

% \subsubsection{Evaluation Metrics}
% We provide a classification report including precision, recall, F1-score, and accuracy for every category in order to assess the model's performance. These measurements provide us a thorough grasp of the model's advantages and disadvantages in terms of identifying malware from benignware. 

\subsubsection{\textbf{Results}}
The CLIP model performed comparatively poorly on the classification challenge. In a binary classification job, the model's accuracy of 0.50 means that half of the data were correctly identified, which is the same as random guessing. The model properly classified half of the samples as Benignware, with a precision of 0.50 for Benignware. The accuracy fell to 0.00 for malware, however, suggesting that the algorithm was unable to accurately detect any malware samples. The model correctly detected all genuine Benignware cases, achieving a flawless recall score of 1.00 for Benignware. However, the malware recall was 0.00, indicating that no malware samples were detected by the model. The F1-score, which weighs recall and accuracy, was 0.00 for malware and 0.67 for benignware. The model performed unevenly in the two classes, as seen by the accuracy, recall, and F1-score macro averages of 0.25, 0.50, and 0.33, respectively. Because the distribution of classes was balanced, the weighted averages of these measures were comparable. %Overall, the CLIP model's performance shows that, while it is good at finding benignware, it has a hard time detecting malicious software. 
This observation implies that the model could not be a good fit for this classification job without large multimodal data, more refinement, or changes.

% \begin{figure}[h]
%     \centering
%     \includegraphics[width=0.75\linewidth]{resources/clip-classification.png}
%     \caption{CLIP classification result}
%     \label{fig:clip result}
% \end{figure}

\section{Discussions}

%Our experimental results aim to address two critical research questions: 

% \begin{tcolorbox}[width=\columnwidth, colback=white, colframe=black, sharp corners=southwest]
%     \paragraph{RQ1}\textit{How effective are various image processing techniques (e.g., RGB vs. grayscale) and their resolutions in distinguishing between benign and malicious Android applications?}
% \end{tcolorbox}
\subsection{RQ1 - Effective Image Attributes}
Diverse image processing approaches significantly affect Android malware detection. RGB setups, especially at 512x512 resolution, offer superior feature representation and accuracy with models, such as EfficientNet-B4 and ResNet-101/152. Grayscale is suitable for resource-constrained scenarios, working well with models such as VGG-16 and MobileNetV2. While high resolutions improve accuracy and ROC AUC, 256x256 balances performance and efficiency, as shown by DenseNet-169 and ResNet-152. Lower resolutions, such as 128x128, suit limited-resource environments, particularly with VGG-16. Ultimately, image processing choices should match task requirements. We provided a model-wise recommendation, which is shown in Table~\ref{tab:optimal_image}.

\begin{table*}[h]
\centering
\caption{Recommended Optimal Image Configurations for Android Malware Classification Models}
\label{tab:optimal_image}
\begin{tabular}{|l|l|c|c|p{7cm}|}
\hline
\textbf{Model}          & \textbf{Optimal Image Set}        & \textbf{Accuracy (\%)} & \textbf{ROC AUC} & \textbf{Comments}                                                                                     \\ \hline
VGG-16                  & Grayscale, 128x128                & 58                    & 0.768            & Offers simplicity and efficiency with relatively balanced performance across all classes.                \\ \hline
VGG-19                  & RGB, 256x256                     & 77.3                  & 0.773            & Optimal resolution and configuration for discriminative performance, avoiding overfitting or noise.     \\ \hline
ResNet-50               & RGB, 512x512                     & 96                    & 0.9906           & Achieves peak performance; RGB 256x256 is a more computationally efficient alternative.                 \\ \hline
ResNet-101              & RGB, 512x512                     & 97                    & 0.9919           & Superior performance at high resolutions; 256x256 provides a computationally efficient option.           \\ \hline
ResNet-152              & RGB, 512x512                     & 96                    & 0.992            & High accuracy and ROC AUC at 512x512; 256x256 offers a balanced trade-off for efficiency.               \\ \hline
DenseNet-169            & RGB, 256x256                     & 96                    & 0.9882           & Best balance between computational efficiency and accuracy.                                            \\ \hline
MobileNetV2             & RGB/Grayscale, 512x512           & 96                    & 0.9903           & Performs equally well in grayscale and RGB at higher resolutions.                                       \\ \hline
EfficientNet-B4         & RGB, 512x512 (Best), Grayscale 256x256 & 97                  & 0.990+           & RGB outperforms grayscale; 512x512 is best for accuracy, while 256x256 grayscale is more efficient.       \\ \hline
\end{tabular}
\end{table*}

%Ultimately, the selection of an image processing approach must correspond to the specific needs of the work. RGB at 512x512 is optimal for achieving maximum accuracy and ROC AUC, especially with high-performance models, but RGB at 256x256 is appropriate for resource-limited environments. Grayscale at 128x128 provides an efficient option for low resource use, particularly with VGG-16. Optimizing the image processing pipeline for the individual model and computing limitations enhances both accuracy and efficiency.  A modelwise recommendation has been demonstrated in Table \ref{tab:optimal_image}

% \begin{tcolorbox}[width=\columnwidth, colback=white, colframe=black, sharp corners=southwest]
%     \paragraph{RQ2}\textit{Does the combination of image and textual data in a multimodal framework improve or diminish the performance of Android malware detection systems?}
% \end{tcolorbox}
\subsection{RQ2 - Multimodal Performance}
The CLIP model, combining image and text, achieves 50\% accuracy but struggles to detect Malware effectively. In contrast, image-only models like ResNet152 deliver higher accuracy and better classification performance, as shown in Table~\ref{tab:clip_comparison}. Given the limited dataset of 34 images and 34 texts, the multimodal approach of CLIP does not offer advantages over optimized image-only models. These results suggest that, under data-constrained conditions, models like ResNet152 remain more effective for Android malware detection.

% The CLIP model's integration of image and textual data does not much exceed the performance of CNN based standalone image-only models. It attains an accuracy of 50\%, although it inadequately identifies Malware. Conversely, the ResNet models, especially ResNet152, provide much more accuracy and enhanced performance in differentiating between Benignware and Malware shown in Table \ref{tab:clip_comparison}.

% This experiment indicates that, given the limited dataset of 34 images and 34 texts, the integration of image and text in CLIP did not enhance the performance of the Android malware detection system relative to conventional image-only models. The findings indicate that, given the limited dataset, the multimodal method may be less successful than the optimally calibrated image-only models such as ResNet152, which exhibit superior accuracy and enhanced classification metrics overall.adjustments or refinement.

\begin{table}[h]
\centering
\caption{Comparative Classification Results for CLIP Model}
\begin{tabular}{@{}lcccc}
\toprule
\textbf{Model}    & \textbf{Accuracy} & \textbf{Precision}& \textbf{Recall}& \textbf{F1-Score}\\ \midrule
\textbf{CLIP}     & 0.50              & 0.25                      & 0.50                    & 0.33                      \\
\textbf{VGG16}    & 0.43              & 0.21                      & 0.50                    & 0.30                      \\
\textbf{VGG19}    & 0.48              & 0.56                      & 0.53                    & 0.43                      \\
 QNN& 0.38& 0.17& 0.38&0.24\\
\textbf{ResNet50} & 0.48              & 0.72                      & 0.54                    & 0.39                      \\
\textbf{ResNet101}& 0.62              & 0.64                      & 0.64                    & 0.62                      \\
\textbf{ResNet152}& 0.71              & 0.71                      & 0.71                    & 0.71                      \\ \bottomrule
\end{tabular}
\label{tab:clip_comparison}
\end{table}

\section{Future work}
We have outlined the research gaps and limitations to guide future research directions:

\begin{itemize}
    \item \textbf{Expand Multimodal Datasets:} Future research should focus on integrating larger and more diverse multimodal datasets (images and text) for Android malware detection. The current study found that the CLIP model, which combines visual and textual inputs, performed suboptimally due to the limited dataset.
    
    \item \textbf{Enhance CLIP with Data Variety:} Improving the CLIP model's effectiveness may involve expanding the dataset to include a broader range of malware types, application descriptions, and images. This could strengthen the model's ability to exploit the connections between textual and visual features.
    
    \item \textbf{Explore Advanced Multimodal Techniques:} Investigating non zero-shot approaches such as attention mechanisms or advanced fusion methods like ViLT (Vision-and-Language Transformer) and VisualBERT could improve the model’s focus on key features from both modalities, potentially boosting classification accuracy.
    
    \item \textbf{Optimize Image-Based Models:} Given the strong performance of ResNet-based models using image data, future work could focus on refining these architectures. This includes exploring deeper or hybrid networks and leveraging advanced transfer learning strategies.
    
    \item \textbf{Evaluate in Real-World Settings:} It is crucial to test these models on large-scale, real-world datasets to assess their scalability, robustness, and effectiveness in detecting diverse malware threats across different environments and platforms.
\end{itemize}

\section{Limitations}
%  Utilizing high-resolution pictures (e.g., 512x512) and intricate models (e.g., ResNet-152, EfficientNet-B4) may be impractical for real-world implementation, particularly in resource-limited settings. Achieving a balance between computing cost and detection accuracy is essential for practical applications. The multimodal technique was assessed using a limited dataset of 34 photos and 34 text annotations, which may not adequately encompass the diverse range of Android malware variants. An expanded and more varied dataset might provide more reliable outcomes and enhance model efficacy.

% An unbalanced distribution of benign and malicious APKs in the dataset may lead to model bias towards the majority class, impairing its capacity to reliably identify the minority class. Mitigating class imbalance via methods such as data augmentation or oversampling is essential for enhancing outcomes.

% Due to the restricted dataset used for training, there exists a danger of overfitting, whereby the models exhibit strong performance on the training data but struggle to generalize to novel instances. Effective regularization methods and cross-validation might alleviate this risk. Again, the used performance measures (accuracy, precision, recall, F1-score, ROC-AUC) are conventional but may not include all dimensions of model efficacy, especially in real-world, dynamic malware contexts where models must adjust to novel and developing threats.

High-resolution images (e.g., 512x512) and complex models (e.g., ResNet-152, EfficientNet-B4) may be impractical in resource-constrained environments, requiring a trade-off between accuracy and efficiency. The multimodal approach was tested on a limited dataset (34 images and texts), which may not reflect the full diversity of Android malware. A larger, more representative dataset could improve reliability and performance.

Class imbalance in the dataset may bias the model toward the majority class, reducing its ability to detect minority cases. Techniques like data augmentation or oversampling are necessary to address this. Additionally, limited training data increases the risk of overfitting, highlighting the need for robust regularization and cross-validation. While standard metrics (accuracy, precision, recall, F1, ROC-AUC) were used, they may not fully capture model performance in dynamic, real-world malware scenarios.

\section{Conclusion}

This study evaluated various image types and resolutions for Android malware detection, finding that high-resolution RGB images combined with models like ResNet-152 and EfficientNet-B4 yield strong performance. We also proposed a multimodal approach using image and text data, with textual features annotated via LLaMA-2. While the CLIP-based model showed promise, it struggled with small datasets, underscoring the need for larger, more diverse data. Future work should focus on improved data integration, advanced fusion methods, and broader datasets to enhance real-world applicability and detection accuracy.

% \section*{Acknowledgments}
% \input{sec_acknowledgments}
 \section{Data Availability}
 The data and materials related to this paper are available at the following link:

\href{https://figshare.com/s/18cb4f07f98d1e6e27ce}{https://figshare.com/s/18cb4f07f98d1e6e27ce}

\bibliographystyle{IEEEtran}
\bibliography{references}  

\end{document}